# Estimation, Analysis and Smoothing of Self-Similar Network Induced Delays in Feedback Control of Nuclear Reactors

Basudev Majumder, Saptarshi Das, Indranil Pan, Sayan Saha, Shantanu Das, Amitava Gupta

*Abstract*—This paper analyzes a nuclear reactor power signal that suffers from network induced random delays in the shared data network while being fed-back to the Reactor Regulating System (RRS). A detailed study is carried out to investigate the self similarity of random delay dynamics due to the network traffic in shared medium. The fractionality or self-similarity in the network induced delay that corrupts the measured power signal coming from Self Powered Neutron Detectors (SPND) is estimated and analyzed. As any fractional order randomness is intrinsically different from conventional Gaussian kind of randomness, these delay dynamics need to be handled efficiently, before reaching the controller within the RRS. An attempt has been made to minimize the effect of the randomness in the reactor power transient data with few classes of smoothing filters. The performance measure of the smoothers with fractional order noise consideration is also investigated into.

## I. INTRODUCTION

IN networked control system (NCS) there is a strong possibility for the packets containing control signal to get delayed because of the shared network medium over which it is being transmitted. These stochastic delays are intrinsically different from the conventional process delays. Process delays are generally large and constant but the network induced delays are stochastically varying and have more adverse effects on the performance of the control system [1]. It has been shown in [2]-[3] that the network induced delay in a Local Area Network (LAN) exhibits self-similarity or non-Gaussian dynamics. The motivation of the present work lies in the fact that if the self similarity of these processes or the delay dynamics associated with the network data can be estimated properly then it is possible to minimize and compensate its deleterious nature. Bhambhani *et al.* [4], Mukhopadhyay *et al.* [5] and Chen [6] have suggested handling fractional order dynamics of network induced delays with fractional order controllers. A robust finite horizon approach to handle FO dynamics [7] in NCS has been proposed by Song *et al.* [8]. Ninness [9] has given several methods for estimating fractional order non-Gaussian $1/f^\alpha$ type noise. But as the fractional order noise and stochastic self-similar network delay are intrinsically different in nature, since the later does not convey any extra energy in the control loop, the FO delay-dynamics in control applications needs to be extensively investigated.

In a nuclear power plant, control signals are generally passed through the dedicated channel from the reactor house to the distantly located control room. But with the advent of cheap communication and off the shelf hardware, Ethernet as a shared medium is getting importance to close real time control loops, in big complex plants. The advantage of using NCS is reduced wiring, modularity and flexibility over the existing technology. Thus online monitoring of reactor data and feedback of control signal can be done easily with LAN. But the major drawback is that congestion occurs in the data transfer processes which leads to delayed control signals resulting in poor control performance. So the data in the control loop suffering from stochastic delays due to the network congestion should not be fed-back directly to the RRS and some filtering or signal processing techniques should be used before it. If these stochastic delays are not compensated the RRS can malfunction and also may lead to tripping of the reactor.

From the reactor physics point of view the dynamics of a nuclear reactor are mainly governed by two different types neutron groups viz. prompt neutron and delayed neutron [10]. At the start up conditions of nuclear reactor or at set-point changes, the random delays are highly detrimental especially for the prompt neutron jump and may cause rapid growth of global power and cause thermal shock to its elements. The present paper applies different types of smoothing filters to the online data measured by the SPNDs to remove random fluctuation in the measured power signal. Also, different configurations of the smoothers are studied with different degrees of self similarity and then the best smoother configuration is reported to handle the network induced delay dynamics. The paper focuses mainly on the following two areas viz. firstly characterizing the delay dynamics of a representative communication channel which can be thought of to be used to feedback the reactor power transient signals and secondly performance study of the smoothing filters to get a jitter-free feedback signal.

The rest of the paper is organized as follows. Section II estimates the degree of self-similarity or Hurst parameter of a representative network data. These random delays have been considered next to corrupt the power transient signal in a nuclear reactor which has been smoothened in section III. The paper ends with conclusion in section IV, followed by the references.

Manuscript received April 14, 2011. This work has been supported by the Department of Science & Technology (DST), Govt. of India under the PURSE programme.

B. Majumder and S. Das is with School of Nuclear Studies & Applications (SNSA), Jadavpur University, Salt-Lake Campus, LB-8, Sector 3, Kolkata-700098, India (E-mail: bmbasudev30@gmail.com).

I. Pan and A. Gupta are with Dept. of Power Engineering, Jadavpur University, Salt-Lake Campus, LB-8, Sector 3, Kolkata-700098, India.

S. Saha is with Dept. of Instrumentation and Electronics Engineering, LB-8, Sector 3, Kolkata-700098, India.

Sh. Das is with Reactor Control Division, Bhabha Atomic Research Centre, Mumbai-400085, India.

## II. ESTIMATION OF SELF SIMILARITY IN REAL COMMUNICATION NETWORK DELAYS

For case study, a heavily loaded LAN data have been collected (Fig. 1) for estimation of the degree of self-similarity indicated by its associated Hurst parameter. Some typical characteristics of such delays in a loaded network [2]-[3] are discussed in the following sub-sections. Fig. 2 shows that the run-time variance of such spiky random variables (packet delays) does not converge to a finite value and have a non-Gaussian $\alpha$-stable distribution [5]-[6].

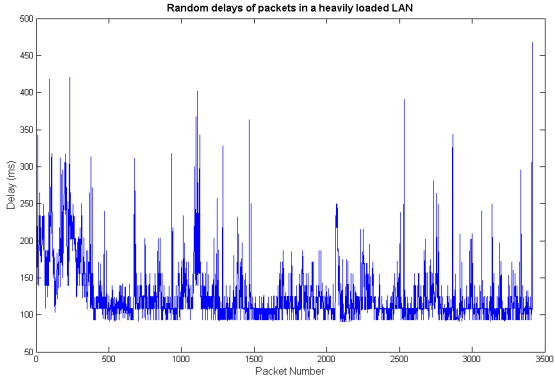

Fig. 1. Time domain presentation of the network induced stochastic delay.

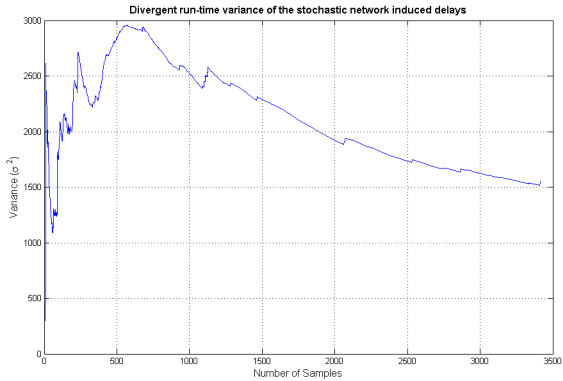

Fig. 2. Diverging run-time variance for the random network induced delay.

### A. Long Range Dependency (LRD)

Let $X_t : t \in \mathbb{N}$ be a time series which is weakly stationary, suggesting that the series has a finite mean and the covariance depends only on the lag between the two points existing in the series. Interaction beyond the two points in the series can result for a strong degree of dependency. The self similarity can be explained in different ways. Let $\rho(k)$ be the auto-correlation function (ACF) of $X_t$. The ACF $\rho(k)$ for a weakly stationary time series $X_t$ is given by: $\rho(k) = E\left[\dfrac{(X_t - \mu)(X_{t+k} - \mu)}{\sigma^2}\right]$ with $E(X_t)$ being the statistical expectation of $X_t$. Also, $\mu$ and $\sigma^2$ are the mean and variance respectively. Self similarity can best be related with long range dependency. The time series $X_t$ is said to have long range dependency if $\sum_{k=-\infty}^{k=\infty} \rho(k)$ diverges. Often $\rho(k)$ takes the form $\rho(k) \simeq C_\rho k^{-\alpha}$, with $C_\rho$ being positive and $\alpha \in (0,1)$. Parameter $\alpha$ is related to the Hurst parameter via the equation $\alpha = 2H - 1$. This is the most common definition of LRD. One very practical view-point of self similarity can be revealed from the signal processing point of view is its power spectral density. If the power spectral density of the time series $X_t$ be $f(\omega)$, then a time-series having LRD must conform to the following relation:

$$f(\omega) = \dfrac{\sigma^2}{2\pi} \sum_{k=-\infty}^{k=\infty} \rho(k) e^{ik\omega} \quad (1)$$

where $i = \sqrt{-1}$. This definition of spectral density comes from Wiener-Khintchine theorem. Also, the weakly stationary time series $X_t$ is said to have LRD if its power spectral density obeys $f(\omega) \simeq C_f |\omega|^{-\beta}$ with $C_f > 0$ and some real $\beta \in (0,1)$ where $\beta$ is related to the associated Hurst parameter by the relation $H = (\beta + 1)/2$.

### B. Hurst Parameter and Its Estimation

Many physical real world processes, exhibit LRD. Modeling of those physical processes require the correct estimation of the LRD which is measured by the Hurst parameter $H$. Estimation of LRD time series enhances the importance of analyzing the self similarity in the time series. The concept of LRD was firstly introduced by Mandelbrot & Van Ness [11] in terms of Fractional Brownian Motion (FBM) [12] and since then it has been addressed by many other contemporary researchers to analyze degree of self-similarity in a time series and leads to the concept of Hurst parameter. A second order time series $Y = f(u)$ is said to have a LRD if its auto-correlation function $\rho(\tau) = E[(f(\tau)f(0)]$ decays with the power law function of lag $\tau$ so that the auto-correlation series $\sum_\tau \rho(\tau)$ is not summable over the length of $\tau$. For the processes having Hurst parameter between $0 < H < 0.5$ are called anti-persistent process or negatively correlated. These processes generally have short range dependency. Processes with $0.5 < H < 1$ are called positively correlated. $H = 0.5$ means the process is not correlated signifying conventional white-Gaussian noise. Processes with $1 < H < 1.5$ is said to have no dependency in time domain.

There are different methods to find out the Hurst parameter of a fractal time series. Popular method to find out the Hurst parameter is the R/S analysis. Apart from that there are a number of methods like aggregated variance method, absolute value method, Periodogram method, variance of residuals method, local whittle method, wavelet based method, Higuchi method and differenced variance approach etc [12]-[16]. Recently, Chen et al. [17] has reported a new Fractional Fourier Transform (FrFT) based Hurst parameter estimator which is more robust than the existing ones. Abrupt shift of mean in the time series or other contaminations can make the series non stationary and result in an over-estimation of Hurst parameter. In the present study, communication data taken from a heavily loaded shared LAN has been analyzed with these estimators as a test case. LRD can be thought of in two different ways. In time domain high degree of correlation in the distant samples of any time series can be modeled as LRD. In frequency domain significant amount of power at very low frequency can be the reason of presence of LRD. Details of few estimators related to the Hurst parameter of a time-series are discussed next.

### C. Rescaled Range (R/S) Analysis

Let $R(n)$ be the range of data aggregated over the block of length $n$ and $S(n)$ be the variance recorded over the same scale of range. For the series to follow self similarity the following relation must be maintained:

$$E[\frac{R}{S(n)}] \simeq C_H n^H \quad (2)$$

Taking logarithm on both sides of (2), the Hurst parameter can be estimated by the following regression formula:

$$\log E[\frac{R}{S(n)}] \simeq \log C_H + H \log n \quad (3)$$

From (3) it is clear that $C_H$ is a positive constant and independent of $n$. Thus, it should have a constant slope as $n$ becomes large. If the process sample is drawn from a stable distribution, slope of (3) will have the value of 0.5 like the random Gaussian noise. If the slope is over 0.5 it indicates the persistency in the time series. If the slope is below 0.5 an ergodic mean reverting process is indicated. Small value of $n$ will make the result anti-persistent and the resulted value of Hurst parameter will be invalid. Again a large value of $n$ will produce too few samples to correctly estimate $H$ for a self similar process. Therefore, choice of $n$ should be judicious with this type of estimator.

### D. Aggregated Variance Method (Aggvar)

Aggregated variance method considers variance of ($X_t^{(m)}$), where $X_t^{(m)}$ is a time series obtained from $X_t$ by aggregating over the $m$ number of blocks:

$$Var(X^{(m)}) \simeq m^{2H-2} \text{ as } \frac{N}{m} \to \infty \text{ \& } m \to \infty \quad (4)$$

### E. Absolute Value Method (Absval)

In the absolute value method, the original time series data $X = X_i (i > 1)$ is divided into blocks of size $m$ and average within each block for successive values of $m$ is given by:

$$X^{(m)}(k) = \frac{1}{m} \sum_{i=(k-1)m+1}^{km} X(i), k = 1, 2, \cdots \quad (5)$$

Then, the data $X_1, \ldots, X_n$ is divided into $N/m$ blocks of size $m$ and the sum of the absolute values of the aggregated series is computed as $\frac{1}{N/m} \sum_{k=1}^{N/m} |X^{(m)}(k)|$. After calculation of the absolute sum, the logarithm of this statistical value is plotted versus the logarithm of $m$. For long range dependent time series with parameter $H$, the slope of the line, thus obtained, is generally $(H-1)$.

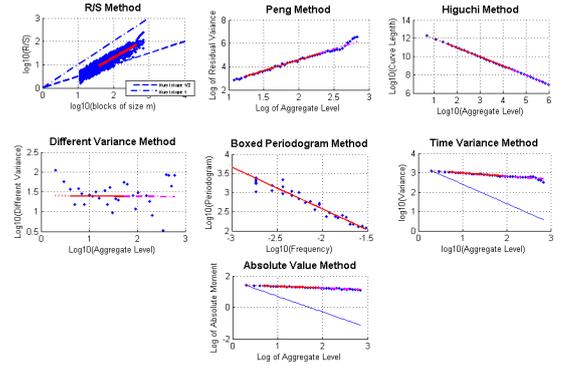

Fig. 3. Estimated Hurst parameter by different statistical techniques.

TABLE I
HURST PARAMETER, FRACTIONAL ORDER AND FRACTAL DIMENSION WITH DIFFERENT ESTIMATORS

| Estimator | Hurst (H) | Fractional Order α=(2H-1) | Fractal Dimension (D=2-H) |
|---|---|---|---|
| Absval | 0.8898 | 0.7796 | 1.1102 |
| Aggvar | 0.917 | 0.834 | 1.083 |
| Boxper | 1.0459 | 1.0918 | 0.9541 |
| Diffvar | 0.9961 | 0.9922 | 1.0039 |
| Higuchi | 0.9987 | 0.9974 | 1.0013 |
| Peng | 0.9346 | 0.8692 | 1.0654 |
| Per | 0.8938 | 0.7876 | 1.2124 |
| R/S | 0.8837 | 0.7674 | 1.2326 |

### F. Periodogram (Per) Method

The periodogram is defined by:

$$I(\xi) = \frac{1}{2\pi N} \left| \sum_{j=1}^{N} X_j e^{ij\xi} \right|^2 \quad (6)$$

where, $\xi$ is the frequency and $i = \sqrt{-1}$. For a series with finite variance, $I(\xi)$ is an estimate of the spectral density of the series. A log-log plot of $I(\xi)$ ought to have a slope of $(1-2H)$ close to the origin.

### G. Differenced Variance (Diffvar) Approach

The series is divided into $n$ groups. Within each partition, the variance, relative to the mean of the total series, is evaluated. The first difference of the variances is then calculated. A measure of the change as a variable parameter of these calculations between different partitions is calculated. The number of groups ($n$) is increased and the same method is repeated. The observed variability changes with increasing $n$ and is related to the Hurst parameter $H$ of the time series, this methodology is known as differenced variance approach. Log-Log plot of variability changes with the number of partition is linear and related with a slope of $H$. Thus, $H$ can be estimated by linear regression.

### H. Higuchi Method

Here, also the series is grouped into $n$ number of partitions. At first the aggregated sums of the series are evaluated. In next step the absolute differences of these cumulative sums between the partitions are analyzed to find out the fractal dimension of the time series. This process is repeated for $n$ number of groups. The result changes as $n$ increase and related to the Hurst parameter $H$ of the time series. A log-log plot of the statistic versus number of partitions is ideally expected to be linear with a slope related to $H$.

*I. Estimation of Hurst Parameter*

Details of other class of Hurst estimators are described in [14]-[16]. Table 1 and Fig. 3 compares the fractional order ($\alpha$), Hurst parameter ($H$) and the corresponding fractal dimension ($D$) of the network induced stochastic delay with different types of estimators. It is seen that the R/S estimator gives minimum value of $H$ indicating LRD in the network induced delay.

### III. STATISTICAL ANALYSIS OF STOCHASTIC DELAYS IN COMMUNICATION NETWORK

*A. Time Domain Analysis*

Few statistical measures [13] are estimated next for the randomly varying network delay. In time domain following factors characterize the time series $E[k]$:

*Mean ($\overline{E}$)*: Mean is generally the average value of the data.

$$\overline{E} = \frac{1}{N} \sum_{k=1}^{N} E[k] \tag{7}$$

*Variance ($\sigma^2$):* Variance is the squared standard deviation and often known as the noise power.

$$\sigma^2 = \frac{1}{N} \sum_{k=1}^{N} (E_n[k])^2 \tag{8}$$

*Skewness ($J$):* Skewness is a non dimensional measure of the symmetry of a distribution. Positive value of skewness shows a long tail in positive direction and a negative value indicates the presence of tail in negative direction.

$$J = \frac{1}{n} \sum_{k=1}^{N} \left( \frac{E_n[k] - \overline{E}}{\sqrt{E_n[k]^2}} \right)^3 \tag{9}$$

*Kurtosis ($k$):* Kurtosis shows whether the data set have sharpness or flatness near its mean. Positive kurtosis mean sharpness or the data sets peak near the mean and negative kurtosis mean that the data sets are flat.

$$k = \frac{1}{n} \sum_{k=1}^{N} \left( \frac{E_n[k] - \overline{E}}{\sqrt{E_n[k]^2}} \right)^4 \tag{10}$$

For the time series representing the random network delays in Fig. 1, the time-domain statistical measures are estimated as the mean $\overline{E} = 127.0536$, variance $\sigma^2 = 9.25$, skewness $J = 2.6617$ and kurtosis $k = 12.9857$.

*B. Frequency Domain Analysis*

In frequency domain due to the existence of the self-similarity, the spectral density falls slowly than the exponential case. Welch power spectral density have been shown in Fig. 4 corresponding to the time-series in Fig. 1 for the analysis of its frequency domain behavior. Chen *et al.* [13] has shown that the spectrum of self similar processes can be better interpreted in fractional domain. Fig. 3 shows significant amount of power in zero frequency.

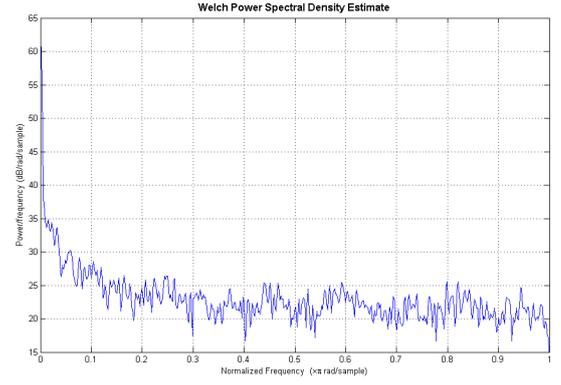

Fig. 4. Welch power spectral density plot for the network induced delay.

### IV. DESIGN OF SMOOTHING FILTERS IN NUCLEAR REACTOR FEEDBACK CONTROL

*A. Different Types of Smoothing Filters*

Measured SPND signal before reaching the regulating system of nuclear reactor gets delayed in the shared communication medium of the networked control loop. The delay dynamics may be Gaussian or may also be of fractional order. Due to the existence of the variable time delay, the packets containing the control signal show spikiness in the time domain which in turn shows that the process is a $\alpha$ stable distribution or obeys fractional order lower order statistics (FLOS) [6]. This measured delayed signal from the simulated shared medium needs to be smoothened in order to compensate the spikiness in time domain. Smoothing can be done with the conventional parametric and non parametric approaches. Moving average (MA) type smoothing filters are most common in parametric smoother family. In non-parametric smoothing technique Savitzky-Golay filter [18] is a popular method to smooth the data. Apart from that there are least square (LS) based filters and Digital smoothing polynomial (DISPO) available for the same purpose. Non-parametric smoothing can also be done with Gaussian kernel regression which tries to fit through the spiky power signal after getting delayed by a random amount at each sample time. All these smoothing filters [19] are expected to perform well to handle white Gaussian noise. But, the behavior and performance of all these smoothers in the presence of Gaussian and fractional-Gaussian stochastic delay has not been explored yet and is the focus of this paper. The above mentioned smoothers are detailed in the following section.

*B. Moving Average Type Smoothers*

A series of raw data $X_1, \ldots, X_n$ is sometimes transformed to a new series of data before it is analyzed, to smooth out local fluctuations in the raw data and this transformation is called data smoothing. This preprocessing is necessary to reduce the effort of the sensors or the actuators. Common type of smoothing filters employs a linear transformation. A linear filter with weights $w_0, w_1, \ldots w_{r-1}$ transforms the given data to weighted averages $\sum_{j=0}^{j=r-1} w_j x(t-j)$ for $t = r, r+1, \ldots, n$. The new data set has length $n - r + 1$. If $\sum_{j=0}^{j=r-1} w_j = 1$ the linear filter is also called an $r$-term moving average filter. If all the weights are equal and they sum to unity, the linear filter is called a simple moving average filter. These smoothers generally take the samples in a variable length of buffer and compute the mean. The size of the window or the variable buffer length can be made proportionally large to get fairly accurate smoothing results. But the restrictions on large buffer size due to the implementation infeasibility and hardware cost motivates us to choose the window size optimally or length of the buffer that produce accurate result with lesser number of flip-flops.

*C. Savitzky Golay Smoother*

Savitzky-Golay smoothers [18] originate directly from particular formulations of the data smoothing phenomena in the time domain. A digital filter is applied to a series of equally spaced data values $f_k = f(t_k)$ with $t_k = t_0 + k\Delta$ for some constant sample spacing $\Delta$ and $k = -2, -1, 0, 1, 2, \cdots$. A simplest type of digital filter replaces the data value $f_k$ by some linear combination of $g_k$ and some number of adjacent members $g_k = \sum_{n=-n_L}^{n_R} c_n f_{k+n}$, with $n_L$ being the number of data points to the left of the data point at $k$ and $n_R$ is the number data points to the right of the data point $k$. An ideal causal filter would have $n_R = 0$. Considering the simplest possible averaging procedure, some fixed values of $n_L = n_R$, $g_k$ is computed as the average of the data points from $f_k - n_L$ to $f_k + n_R$. This is called the moving window averaging with constant $c_n = 1/(n_L + n_R + 1)$ similar to the case above. If the underlying function is considered to be constant or changes linearly with time, then there is no bias present in the result. The moving window averaging does preserve the zeroth moment which is the area under a spectral line. It also preserves its first moment which is the mean position in the buffer size ranging from $-n_L$ to $n_R$ in time domain. What is violated is the second moment, equivalent to the line width. The idea of Savitzky-Golay filtering is to find filter coefficients $c_n$ that preserve higher moments. Equivalently, the idea is to modify the underlying function within the moving window not by a proportional constant but by a polynomial of higher order. Savitzky-Golay filtering manages to provide smoothing without loss of resolution. It assumes that relatively distant data points have some strong redundancy that can be used to reduce the level of randomness in the form of Gaussian or non-Gaussian type stochastic delay.

*D. Kernel Smoothing (Non Parametric Approach)*

A kernel smoother is a statistical technique for estimating a real valued function $f(x): x \in \mathbb{R}^p$ by observing the noisy samples at its input. This has been adopted in our case to examine if it gives any further improvement to the spiky nature of the delayed SPND data containing the reactor power signal. The estimated function is smooth and the level of smoothness is set by a parameter called the bandwidth. Actually, the kernel smoother represents the set of irregular data points as a smooth line or surface. Let $k_{h_\lambda}(X_0, X)$ be a kernel defined by:

$$k_{h_\lambda}(X_0, X) = D\left(\frac{\|X - X_0\|}{h_\lambda(X_0)}\right) \quad (11)$$

where $\{X, X_0\} \in \mathbb{R}^p$, $\|\cdot\|$ is Euclidean norm, $h_\lambda(X_0)$ is a parameter known as kernel radius, $D(t)$ is a positive real valued function whose value is not increasing between $X$ and $X_0$. Popularly known kernels are Epanechnikov, Tricube and Gaussian. We have used Gaussian kernel in the present nonparametric smoothing problem. In the present simulation, the bandwidth parameter is chosen in such a way so that the probability mass function around a point is not wide enough to bring about the under or over-smoothing because the parameter bandwidth controls smoothness or roughness of a density estimate.

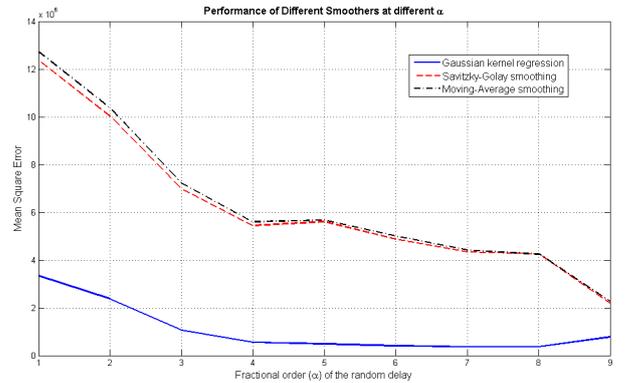

Fig. 5. Performance of different types of smoothers in fractional order domain.

Fig 5 shows the performance of different smoothers with different fractional order of the delay dynamics. It is interesting to note that the smoothers work well near order one. In other words, as the randomness is going from fractional order to Gaussian nature, smoothers are performing well. Fig. 5 also reveals that the nonparametric approach of smoothing outperforms the Savitzky-Golay and Moving-Average smoothing techniques. From Fig. 4 we can infer that at the estimated order of the random delay over LAN based control loop data should be smoothed by Gaussian kernel based smoothing technique, since it gives the minimum Mean Squared Error (MSE) at the statistically determined fractional order of the network induced delay.

*E. Application in Networked Nuclear Reactor Control*

The performance of these smoothing filters are tested for a step-change in reactivity in a nuclear reactor model governed the point-kinetic equations [10] given by (12)-(13) with $\{P, C, \rho\}$ being the reactor power, delayed neutron precursor concentration and reactivity respectively.

$$\frac{dP(t)}{dt} = \frac{\rho(t)-\beta}{l}P(t) + \lambda C(t) \qquad (12)$$

$$\frac{dC(t)}{dt} = \frac{\beta}{l}P(t) - \lambda C(t) \qquad (13)$$

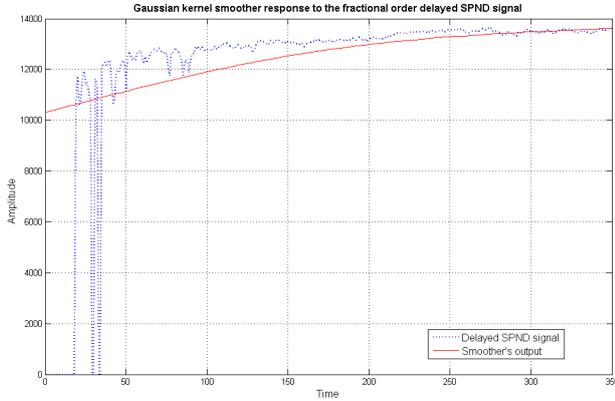

Fig. 6. Performance of kernel smoother over the delayed SPND data.

The measured power signals by the SPND are considered to be transmitted through the LAN from the reactor house to the control room or RRS. But the network induced random delays can make the system malfunction and lead to tripping of the RRS. The corrupted power signal needs to be smoothened before reaching to the controller. Randomness in the form of fractional order delay, introduced by the network deteriorates the transient dynamics of the nuclear reactor, especially at the early stages governed by the prompt jump of the fast neutrons which is generally the output of nuclear fission process. Reactor dynamics due to the delayed neutrons suffer less compared to that with the prompt neutrons. Kernel smoother handles both the condition efficiently to make the control system jerk-less as shown in Fig. 6 where the initial offset is caused by the prompt jump of the fast neutrons.

## V. CONCLUSION

Degree of self-similarity in random delay dynamics over LAN is estimated. A corrupted nuclear reactor power signal is smoothened with few classes of smoothing filters that gives jerk-less signals for reactor control purpose. More involved theoretical analysis for the random network delay dynamics [21] can be investigated in future research.